# Learning about SANS Instruments and Data Reduction from Round Robin Measurements on Samples of Polystyrene Latex

Adrian R. Rennie,[a*] Maja S. Hellsing,[a] Kathleen Wood,[b] Elliot P. Gilbert,[b] Lionel Porcar,[c] Ralf Schweins,[c] Charles D. Dewhurst,[c] Peter Lindner,[c] Richard K. Heenan,[d] Sarah E. Rogers,[d] Paul D. Butler,[e] Jeffery R. Krzywon,[e] Ron E. Ghosh,[f] Andrew J. Jackson[g] and Marc Malfois[h]

[a] *Materials Physics, Uppsala University, Box 516, SE-75120 Uppsala, Sweden,* [b]*Bragg Institute, Australian Nuclear Science and Technology Organisation, Locked Bag 2001, Kirrawee DC, NSW 2232, Australia,* [c]*Institut Laue Langevin, 6 rue Jules Horowitz, F-38042 Grenoble Cedex 9, France,* [d]*ISIS Facility, Rutherford Appleton Laboratory, Didcot, OX11 0QX, UK,* [e]*NIST Center for Neutron Research, 100 Bureau Drive, MS 6100, Gaithersburg, MD 20899-6100, USA,* [f]*Department of Chemistry, University College London, 20 Gordon Street, London WC1H 0AJ, UK,* [g]*European Spallation Source ESS AB, P.O Box 176, SE-221 00 Lund, Sweden,* and [h]*Diamond Light Source, Harwell Science and Innovation Campus, Didcot, Oxon., OX11 0DE, UK. E-mail: Adrian.Rennie@physics.uu.se*

**Synopsis**  Round-robin measurements of polystyrene latex samples at a number of small-angle scattering facilities have been used to assess the current limits of reliability and reproducibility of scattering data, as well as the accuracy of parameters derived from subsequent analysis such as the radius and polydispersity. These results are used to develop understanding and improve the technique.

**Abstract** Measurements of a well-characterised 'standard' sample can verify the performance of an instrument. Typically, small-angle neutron scattering instruments are used to investigate a wide range of samples and may often be used in a number of configurations. Appropriate 'standard' samples are useful to test different aspects of the performance of hardware as well as that of the data reduction and analysis software. Measurements on a number of instruments with different intrinsic characteristics and designs in a round robin can not only better characterise the performance for a wider range of conditions but also, perhaps more importantly, reveal the limits of the current state of the art of small-angle scattering. The exercise, followed by detailed analysis, tests the limits of current understanding as well as uncovers often forgotten assumptions, simplifications and approximations that underpin the current practice of the technique. This paper describes measurements of polystyrene latex, radius 720 Å with a number of instruments. Scattering from monodisperse, uniform spherical particles is simple to calculate and displays sharp minima. Such data test the calibrations of intensity, wavelength and resolution as well as the detector response. Smoothing due to resolution, multiple scattering and polydispersity has been determined. Sources of uncertainty are often related to systematic deviations and calibrations rather than random counting errors. The study has prompted development of software to treat modest multiple scattering and to better model the instrument resolution. These measurements also allow checks of data reduction algorithms and have identified how they can be improved. The reproducibility and the reliability of instruments and the accuracy of parameters derived from the data are described.

**1. Introduction**

When using any experimental method for measurement of structure or properties of materials, it is important to understand the true uncertainties associated with the data and the derived results from the analysis of the data. General scientific conclusions should be independent of details of particular measurement methods or instruments. Identification of anomalies between measured data can be a valuable route to providing better understanding. In this respect, comparison of measurements from identical samples that are made using different instruments and a range of complementary techniques is helpful. Modern small-angle neutron scattering (SANS) instruments at high flux facilities allow collection of data with excellent statistical accuracy. One may hope that analysis of the data could match this accuracy. However, this imposes generally a requirement that the calibrations of instruments and modelling of the scattering would have similar precision. In some circumstances, only the relative differences between samples are important and so factors that limit the absolute calibration may be of less significance. SANS instruments are available only at a relatively small number of facilities. The limits on available measurement time restrict the number of control measurements that can be performed and the repeated collection of data for similar samples during studies of specific materials. This makes it particularly important to document as well as possible the different uncertainly limits of the data. Although SANS measurements are of particular value when exploiting the contrast opportunities that are available using isotopic substitution, particularly of hydrogen and deuterium, or with polarised neutrons, the desire to make a comparison with other techniques has led us to use a simple system of spherical particles dispersed in heavy water ($D_2O$) in a round-robin study.

Understanding the relative and absolute uncertainty limits (i.e. accuracy) of SANS data is important for a range of problems and for the interpretation of results. Trewhella (2008) has argued for better reporting of the results in the area of small-angle scattering from biological macromolecules. This has led to proposals for specific standards of presentation of small-angle X-ray and neutron scattering data from these materials in publications (Jacques et al., 2012). Measurement and comparison of known samples is a route to obtaining the necessary understanding of the limits of the techniques. One single sample cannot readily test all aspects of the performance of a small-angle scattering instrument. Calibration of intensity, momentum transfer, uniformity of detector response and resolution are each important. Materials that provide strong scattering signals in one range of momentum transfer may not be appropriate for other configurations. For example, in order to test for uniformity of a detector, it may be desirable to have a sample that has scattering that is isotropic and does not vary strongly with momentum transfer. In contrast, to assess the effects of instrument resolution sharp maxima and minima in the scattering are helpful. Ideal test samples will probe the limits of current assumptions and approximations, thus highlighting if and when the usual approximations break down and what their effects are on data interpretation.

We report data obtained at several facilities for scattering from samples of spherical particles that had been prepared for colloid physics experiments (Hellsing et al, 2012) but fortuitously proved convenient for this comparative study of small-angle scattering measurements. Polystyrene latex is readily synthesised by emulsion polymerisation. When the surfactant is removed from the product by dialysis, the particles are stable with respect to flocculation because of ionisation of surface groups that remain from the ionic initiator used for the polymerisation. These particles are relatively monodisperse and of uniform density. The simple scattering from uniform spheres makes comparison of measured data with calculated models straightforward.

A number of sample materials have been used as secondary standards to verify calibration of small-angle scattering instruments. For example, semi-crystalline polyethylene ('Lupolen') has been widely used on small-angle X-ray scattering instruments. As this is not very stable, particularly when exposed to intense beams of synchrotron radiation, other materials have been proposed more recently such as glassy carbon (Zhang et al., 2010). Neither of these materials has scattering patterns that are easy to model and so understanding the differences that have been observed is not straightforward.

## 2. Materials and Measurements

### 2.1. Samples

Polystyrene latex (designated PS3) was prepared following the procedure that has been described previously by Goodwin et al. (1974). Styrene (73.4 g, Merck ≥ 99%) was distilled at low pressure to remove inhibitor. An emulsion polymerisation was carried out in deionised water (718 g), under nitrogen with added sodium dodecyl sulfate (1.01 g, Sigma Aldrich ≥ 99%). Potassium persulfate was used as the initiator (0.5 g, Fluka ≥ 98%) and the reaction was allowed to continue for 24 hours at 60 ºC with constant stirring. The latex was then dialysed extensively (water resistivity 18 MΩ cm) to remove ionic impurities and surfactant residues that remained from the polymerisation. The surface $\zeta$-potential was determined as −31 mV using a Malvern Zetasizer nano. The polystyrene latex consists of uniform particles with a density of 1.05 g cm$^{-3}$. The coherent scattering lengths for neutrons for the elements and individual isotopes are well documented (Sears, 1992). The neutron coherent scattering length density of the particles with an empirical formula $C_8H_8$ is calculated as $1.41 \times 10^{-6}$ Å$^{-2}$.

The concentration of the latex dispersion was determined by drying a small amount of the cleaned sample to constant weight. In order to reduce the correlation between the latex particles that would exist in deionised water measurements were made in the presence of added electrolyte and thus provide simpler scattering data. A stock solution of 1 mmol L$^{-1}$ NaCl in $D_2O$ was prepared and a series of samples identified as A, B and C with the composition shown in Table 1 were obtained by dilution of the latex with the salt solution. Sample A was prepared by diluting the concentrated latex stock and other samples by successive dilution. For light scattering further dilutions were made with the stock salt solution to reduce multiple scattering. $D_2O$ was used as the dispersion medium to minimise the incoherent background in neutron scattering measurements. The isotopic purity was 99.7% D and after dilution of the original stock of latex the final isotopic compositions were as shown in Table 1.

For the light scattering measurements, further dilutions were prepared that also used 1 mmol L$^{-1}$ NaCl in D$_2$O. The first dilution had a particle mass fraction of $4\times10^{-5}$ and a subsequent dilution provided about $3\times10^{-6}$ mass fraction. All samples, as well as the NaCl solution in D$_2$O used as the dispersion medium, were filtered through Millipore 0.22 μm PVDF filters.

**2.2. Neutron Scattering Instruments**

Small-angle neutron scattering measurements were made at four different facilities. The study used five instruments: one, SANS2D, uses a pulsed 'white' beam and the time-of-flight method to determine the neutron wavelength, λ. Other instruments all used mechanical velocity selectors to provide a continuous, approximately monochromatic neutron beam. A summary of the different instruments and wavelength resolution is provided in Table 2. In all cases, the collimation of the incident beam is provided by apertures that could be selected together with the effective source distance that is altered by inserting neutron guides to provide an appropriate beam divergence for each measurement configuration. Data were recorded on two-dimensional detectors.

The measured data were corrected using software provided at each facility for background scattering making allowance for the measured sample transmission, detector uniformity and instrument noise. The data were converted to one-dimensional sets of intensity versus the momentum transfer, $Q$ by averaging over the different azimuthal angles on the detector and choosing appropriate radial bins. $Q$ is calculated as

$$Q = (4\pi/\lambda) \sin(\theta/2) \qquad (1)$$

where λ is the neutron wavelength and θ is the scattering angle. Further details of the components such as detectors as well as the data reduction software can be found in the references cited in Table 2. As far as possible, the data were exported from the reduction software in the canSAS 1-D data format (http://www.cansas.org/formats/canSAS1d/1.1/doc/overview.htm) or a conversion program was used so that they were accessible to a broad range of analysis software.

The theoretical model for scattering from uniform spheres is straightforward. The scattering for a sphere of radius R, is described by a form factor, P($Q$)

$$P(Q) = [3\{\sin(QR) - QR \cos(QR)\}/(QR)^3]^2. \qquad (2)$$

The form factor is normalised to 1 at $Q$ equal to zero but the measured intensity of scattering extrapolated to $Q = 0$ will be given by a scale factor

$$I(Q=0) = n \, V^2 \, (\Delta\rho)^2 \qquad (3)$$

where $n$ is the number density of spherical particles of volume V and $\Delta\rho$ is the difference between the scattering length density of the particles and that of the dispersion medium.

### 2.3. Characterisation with Other Techniques

#### 2.3.1. Microscopy

The particles were inspected using both atomic force microscopy and scanning electron microscopy. A Nanosurf Mobile S was used to make a topographic scan in tapping mode of particles that had been allowed to dry on a silicon wafer. A sample that had been diluted more was examined using a scanning electron microscope (FEI Model XL30 operated at 5 keV) without coating the particles. These images are shown in Figure 1. The microscope images show that the particles are spherical and have mean diameters of about 1450 Å but the resolution does not allow a precise determination of the size distribution. Higher resolution images could be obtained using transmission electron microscopy but the coating required to obtain high quality micrographs can significantly perturb the observed size. For this reason, quantitative analysis was not pursued further.

#### 2.3.2. Static and Dynamic Light Scattering

Light scattering measurements have been performed in the Partnership for Soft Condensed Matter laboratory at the ILL, Grenoble using an ALV CGS-3 DLS/SLS Laser Light Scattering Goniometer System (ALV GmbH Langen, Germany). This instrument allows for simultaneous measurement of static and dynamic light scattering for the range of scattering angles $25° < \theta < 155°$. It is equipped with a helium-neon laser operating at a wavelength (in vacuum) of 6330 Å and a power of 22 mW. The ALV/LSE-5004 electronics is used with an ALV-7004 fast multiple tau digital correlator. Scattering data were recorded using a pseudo-cross correlation arrangement, consisting of a fibre-optical detection unit with a fibre based beam splitter and two avalanche photodiode detectors. Toluene was used as calibrant for the intensity. Static light scattering can be analysed in a manner that is largely analogous to small-angle scattering data. The values of momentum transfer must make allowance for the refractive index of the sample, which being dilute, is approximately that of water (1.33). The long wavelength provides data at small $Q$ and so a Guinier analysis with a straight-line fit in a plot of ln (Intensity) versus $Q^2$ can readily provide the radius of gyration, $R_g$. Example data with the fit are shown in Figure 2(a). Error bars indicate ± 1 standard deviation. The derived radius of gyration is 555 Å for the sample with mass fraction $2.6 \times 10^{-6}$ and this corresponds to a sphere of Z-average radius $\sqrt{(5/3)}R_g$ or 717 Å. The statistical uncertainty (standard error) in the regression coefficient is better than 0.5%. For the data measured at the higher concentration, the derived particle radius is 724 Å. These values are in good agreement with the microscopy results.

Dynamic light scattering is used to determine the spectrum of relaxation times within a sample and the correlation time $\tau$ is related to the translational diffusion coefficient $D_T$ by:

$$D_T = 1 / (Q^2 \tau) \qquad (4)$$

in the case of dilute dispersions. The hydrodynamic radius, $R_H$, of the particle is related to $D_T$ and the thermal energy by the Stokes-Einstein equation

$$R_H = k_B T / (6 \pi \eta D_T) \qquad (5)$$

where $k_B$ is Boltzmann's constant, T is the absolute temperature and η is the dynamic viscosity. In these calculations it is important to use the viscosity of $D_2O$ which is 1094 μPa s at 25 °C (Kestin et al., 1985) and is about 20% higher than that for $H_2O$. The change in viscosity of water with 1 mmol $L^{-1}$ NaCl is negligible compared with the other uncertainties (Kestin et al., 1981; Zhang & Han, 1996). An example of the light scattering data measured at a scattering angle of 90° is shown in Figure 2(b) with a fitted correlation function. The hydrodynamic radii derived from a cumulants analysis of the data for the two samples with mass fractions $4\times10^{-5}$ and $3\times10^{-6}$ were 720 and 728 Å respectively. There was insignificant variation observed in $D_T$ with scattering angle as expected for monodisperse spheres. The dominant experimental uncertainty in the radius derived from the light scattering is likely to arise from possible variation in the temperature as causes a large change of viscosity. An uncertainty of about 0.2 to 0.4 °C gives rise to errors of less than 1%. The radius was estimated to have a distribution with a standard deviation of about 35 Å although the finite number of time intervals on the correlator limits the precision significantly and may cause an overestimate of the width of the distribution. It is important to note that the good direct numerical agreement between the hydrodynamic radius and the Z-average radius from the Guinier analysis would not be expected unless the particles were effectively hard, non-interacting spheres. The ratio of the radii obtained by the two techniques is sometimes considered as a shape factor that provides information about anisotropy or non-uniformity of particles.

### 2.3.3. Small-angle X-ray Scattering

For purposes of comparison with neutron scattering, the samples prepared for SANS experiments have also been measured using X-rays (SAXS). Data were collected at the Diamond Light Source beam line I22 using an energy of 12.4 keV (equivalent to 1.00 Å wavelength) and sample to detector distance of 9.2 m. Data were recorded using a Pilatus 2M detector giving data in a range of momentum transfer from 0.0064 $Å^{-1}$ to 0.169 $Å^{-1}$.

### 3. Small-Angle Scattering Results and Interpretation

Data measured for Sample A with each of the SANS instruments included in the study is shown in Figure 3. The different characteristics of the instruments, particularly the ranges of momentum transfer and the resolution, in the configurations that were used are apparent from inspection of this data. For example, on time-of-flight SANS instruments such as SANS2D, data can be measured with comparatively good resolution in d$Q$/$Q$ over a wide range of $Q$ in a single configuration. This eases the analysis of the data and other, different aspects of the experiment that may cause smoothing of the measured data such as polydispersity of size or multiple scattering can be identified more readily.

It is difficult to assess the detailed differences between the data sets that are shown in Figure 3 and alternative plots are helpful. Simply expanding the scales in selected regions such as that shown in Figure 4(a) allows the systematic differences between the data sets to be seen more clearly. These are most marked at low values of $Q$ and around the minima in the form factor P($Q$). Although at first glance there is good agreement between measurements shown in Figure 3, the differences of about 15 to 20% in reported intensity at a given mean Q value seen in Figure 4(b) are much larger than the uncertainty bars that show ± 1 standard deviation. The derived intensity scale for each data set used procedures that vary between the instruments and depend on the design (some further details are available at: http://www.cansas.org/wgwiki/index.php/Calibration_Procedures). For the monochromatic SANS instruments, measurements of the direct beam using calibrated attenuators are most common but SANS2D, because of the special challenges of measurements with a broad wavelength band, uses the intensity of scattering from secondary polymer standards as described by Wignall and Bates (1987) who assessed the standard uncertainty of such methods as about 6%. A similar procedure with secondary standards was also used on D11. Most data reduction packages estimate the uncertainty in the intensity at a given momentum transfer either from Poisson counting statistics or from the distribution of values of intensity in individual pixels that are averaged into a single $Q$ bin or point. It should be emphasised at the outset that these differences, even assuming that the samples were identical, do not necessarily imply that any given data set is incorrect but rather that the description of the data may be incomplete. The experimentalist is primarily interested in the information that can be deduced about samples from a measurement rather than the data itself. It is therefore useful to turn attention from simple inspection of data to consider how it may be interpreted.

To obtain good fits of a model to the data, three terms that broaden the scattering function from monodisperse spheres were included. These involved instrument resolution, polydispersity and multiple scattering. In principle, the resolution for each $Q$, which was assumed to be a Gaussian function with a width that changes as $Q$ increases (Pedersen et al., 1990), is determined from instrument geometry and calibration of the wavelength spread. The results of the model fit are shown in Figure 5 for the data measured on D22 for the three concentrations. When data are recorded over a sufficient range that many minima are visible, it is clear that for the most concentrated sample, broadening only with polydispersity and resolution to fit the low $Q$ region would smear too much the data at high $Q$ as shown in Figure 6. This problem is not apparent for the lowest concentration and so the different contributions to broadening can be distinguished in a simultaneous fit to the different data sets. The high statistical quality data from a number of instruments has prompted development of a simple algorithm for fitting that can include double scattering, polydispersity and resolution (Ghosh and Rennie, 2012). Allowing for the smearing due to double scattering, and constraining the size and polydispersity to be identical for each concentration, gave a mean radius of 724±3 Å and a standard deviation of a Gaussian size distribution of 20±5 Å. The fraction of beam that was scattered by the sample was fitted as 4, 2 and 1% for the samples, A, B and C respectively. If the effects of double scattering are not included in the model fitting and only resolution and polydispersity are used to smear the model, the parameters change to a mean radius of 727±3 Å and a standard deviation of the size distribution of 32±4 Å. The fit is noticeably poor in not smearing sufficiently the first minimum for sample A and smearing too much the model at larger values of $Q$. If only data for sample A (the most concentrated) are modelled with resolution smearing but no double scattering, the polydispersity increases to 38±4 Å. If data for a given instrument are only available in a limited range of $Q$, it is important to constrain the polydispersity to that found from other measurements to ensure good model fits with reasonable parameters for the resolution. The similarity of the influences of these factors in smearing the data emphasizes the need for comparative measurements. Comparisons and model fits of measurements with different concentrations of particles as shown in Figure 5 or different sample thicknesses are the simplest approach to quantitative identification of the effects of multiple scattering.

The uncertainties that are obtained from minimisation in a fit cannot be taken simply as the total uncertainty in the resulting parameter unless it is established that all scattering and instrumental effects are included correctly. The systematic effects of double scattering could alter the estimate of polydispersity by a factor of two! Similarly, uncertainty in the resolution in wavelength or in the angular spread of the beam could give rise to significant differences. In this respect, changes in $\Delta\lambda/\lambda$ that are correlated with the instrument collimation that defines the angular divergence after the velocity selector are poorly documented. Relatively small differences can make large changes to some parameters. In the case of the measurements on latex, the sensitive parameter is the polydispersity but in other studies, anisotropy of scattering objects or models of non-uniform density might be altered. If data are available only for a restricted range of $Q$, then modelling may be more difficult and even the values for the particle size can be modified if the relative positions of several sharp minima are not available to constrain the fit.

Dividing one data set by another measured with a different particle concentration under identical conditions provides a good means to identify differences that do not depend on any features of the instrument or data reduction. Any effects of variation in detector efficiency or solid angle as well as absolute scaling and instrument resolution should not appear in the ratio. A plot of data measured with two different concentrations of permanently formed particles or polymers will identify the effects of multiple scattering and any possible interactions that change with concentration. Such a plot of the ratio of scattering from sample C to that of sample A is shown in Figure 7. The average corresponds reasonably with the ratio of concentrations. The sharper minima in the scattering from sample C that has less multiple scattering are apparent as dips in the curve. At large values of $Q$ there are some systematic deviations that arise from the different level of incoherent scattering. Interactions between particles would be apparent at low $Q$ but no significant effects of a structure factor changing significantly from unity are seen. Monte Carlo modelling of non-interacting polydisperse spheres with radius 720 Å is also shown in Figure 7. This modelling used the NIST IGOR Pro SANS package (Kline, 2006) that uses an analytic function for the scattering from a sphere. Only modest approximations to the D22 instrument configuration as regards resolution could be made with the program but the data were modelled well: the significant test of multiple scattering is that the ratio of the measurements at different concentrations seen in the experimental data is observed in the simulation. Treatment of multiple scattering has been described by a number of authors. For example, Schelten and Schmatz (1980) indicate how multiple convolutions and theory of Fourier transforms can be exploited to calculate the effect. Simulations such as that described by Copley (1988) treat the case of spherical particles specifically and show similar phenomenology to those effects seen in the simple convolution model and the Monte Carlo simulation in Figure 7. A single extra convolution integral in a fitting program can include the effect of double scattering and thus most of the multiple scattering unless the transmission is small. Although this is only an approximation to the full multiple scattering, programs can readily incorporate this along with other effects of resolution and polydispersity (Ghosh &

Rennie, 2012). The different fits to a single data set shown in Figure 6 indicate why it is necessary to include both the influence of polydispersity and double scattering, as well as resolution to model the data. The overall reproducibility of the intensity is best judged by the value of the intensity extrapolated to $Q$ equal to zero or the fitted contrast. The scatter in the values for this intensity is about 10%.

Detailed descriptions of how different factors and constraints alter fits to the data from the complete range of instruments are not of general interest but a brief summary of the extent of the agreement of the various results obtained by model fitting is useful. Data from different instruments gave fitted values for the mean radius that varied between about 708 and 735 Å with fitting uncertainty of about ±5 Å. The optimisation of the fits for polydispersity was more difficult and values in the range 15 to 30 Å were obtained with uncertainties in the minimisation of about 8 to 10 Å but these are highly dependent on the assumptions made about the resolution. Apart from fitting models, other methods of analysis are sometimes used to interpret small-angle scattering data. These include indirect Fourier transforms of the data to obtain distributions of distance, and even simple analysis of straight line fits such as Guinier plots and evaluation of integrals like the scattering invariant (see e.g. Brumberger, 1995). However, it is more difficult to include the effects of resolution and multiple scattering in such analysis and so such comparisons are not helpful.

The small-angle X-ray scattering (SAXS) data provide an interesting comparison as the electron density difference between water ($D_2O$) and polystyrene is rather small and corresponds approximately to a scattering length density difference of $0.2 \times 10^{-6}$ Å$^{-2}$. Data are shown in Figure 8 for sample A. The lowest $Q$ is about 0.0065 Å$^{-1}$. The small signal reaches the background scattering at $Q$ of 0.08 Å$^{-1}$. Although the data does not extend over a very wide range, the visibility of 13 minima gives a clear indication of the particle size. The Porod plot indicates that the scattering is not decreasing as $Q^{-4}$ but rather oscillates about a slope of $Q^{-3}$ that suggests a surface layer of different density may be present. Even for this X-ray data, it is clear that detailed modelling must take account of the instrument resolution that is dominated by the horizontal divergence of about 50 μrads as well as the finite size of both the beam and the detector pixels. A model with a sphere and a concentric shell suggests a particle size with a radius of about 720 Å with the outer region of approximately 20 Å having a higher electron density by about 13% that is consistent with the presence of the initiator residues that form ionisable groups on the surface. The sulfate groups and sodium counter ions have a higher scattering length density than polystyrene and diminish the contrast for neutrons of the particles with respect to $D_2O$. The contribution to SANS intensity from such a shell would be about 500 times less than that from the core of the particles at low $Q$ of about 0.005 Å$^{-1}$ and would not be visible in the data. For SAXS the scattering from the shell can be dominant as the difference in the electron density that provides contrast between polystyrene and water is small ($\Delta\rho_X$ is less than $0.2 \times 10^{-6}$ Å$^{-2}$).

## 4. Discussion and Conclusions

### 4.1. Lessons that are learnt from the Round-Robin Measurements

Use of the same samples on a range of instruments has allowed the comparability and reproducibility of data to be tested. A first conclusion is that simply comparing measured intensity at a particular mean scattering vector without allowance for the effects of resolution is not helpful. The large differences seen in Figure 4 are primarily due to resolution. This idea is important as it demonstrates that the temptation to arbitrarily scale, as regards intensity, data obtained with different instrument configurations so that a few points 'match' in regions of $Q$ that overlap could significantly distort the shape of a resulting combined data set. The benefits of placing all data independently on an absolute scale of intensity are thus highlighted. Although calibration procedures for a given instrument may generate data that are reproducible to about 1 to 3% the absolute accuracy depends on a range of factors that makes the uncertainty higher and often standard uncertainties of 5 to 10% are estimated.

Measurements of samples with a well-known form factor, such as spherical particles have allowed a number of specific problems with procedures for data reduction and analysis to be identified at facilities. Treating data with a variety of software packages has highlighted some deficiencies in metadata. In a number of cases, even the raw data files did not contain adequate details to calculate the resolution. Similarly, parameters to determine absolute intensity or to scale the data were missing. Facilities are working to rectify these difficulties. In some cases, the software estimated the smearing due to resolution to be significantly more than that observed in the data and corrections were needed to either the metadata or algorithms to provide reasonable results. We conclude that it is clearly helpful to measure samples with known, sharp features in the scattering to test such calculations particularly as the calculated resolution may have large influences on some derived structural parameters such as polydispersity.

As the calculations of instrumental resolution have improved, the need to have accurate values of the wavelength distribution from velocity selectors, which can vary with collimation, has been identified. Data from time-of-flight measurements at pulsed sources are often recorded with sufficiently good time resolution that the choice of final $Q$-bin sizes can be made after the experiment to optimise the reduced data with respect to either statistics or resolution. Incorporation of the calculations in the primary data reduction package is an essential prerequisite for further progress in analysis of SANS data. The studies have highlighted how even modest amounts of multiple scattering can significantly alter measured data and the results have prompted development of efficient fitting algorithms to include a simple approximation that includes double scattering.

In general, the spread in the fitted mean particle size is about ±1% but the uncertainty in deducing the distribution of size is very much larger and requires careful understanding of a number of effects. At low $Q$, when the intensity changes rapidly with $Q$, the differences observed for the intensity depend significantly on resolution. It was not particularly the aim of this work to provide a further secondary standard for repeated use on scattering instruments but latex samples could, if carefully sealed and stored, be used for this purpose. Care is necessary to avoid aggregation and even sedimentation or creaming due to density differences. The latter problem could be mitigated by tumbling the sample, or could be reduced by density matching the particles and the dispersion medium.

The different scattering observed in the SANS and SAXS studies makes the direct comparison more difficult but demonstrates the value of combining such data. It would be possible to prepare samples that had higher contrast for X-rays and so gave less signal from structure at the surface of the particles, for example by using mixtures of ethanol and water as the dispersion medium. One could even density match the particles in this mixture using $D_2O$ with either normal or perdeuterated ethanol so as to avoid problems of sedimentation or creaming during storage of samples.

## 4.2. Suggestions and Future Work

The uncertainty in fitted parameters, such as radius and polydispersity, is limited less by statistical uncertainty in data and fitting but more by knowledge of systematic errors in calibration and modelling. For a given data set and analysis procedure, the uncertainty (standard deviation) in radius, for example, may be just 0.5 to 1% but the spread of values fitted to a single data set under different assumptions can be 2% or more. Polydispersity may be affected by as much as 50% if incorrect assumptions are made. This scatter in sample parameters derived from data depends mostly on the choice of what factors are included in the analysis and the correct incorporation of calibration constants and other instrument parameters. The recognition that even in neutron scattering experiments, random errors from counting statistics are often not dominant is valuable. Relative differences between samples may be determined to higher precision than absolute values of parameters.

These conclusions about uncertainty have provided input to proposals for future reduced data formats (Jemian et al., 2012) that should have the capability to better document the different sources of error and potentially allow absolute and relative errors to be incorporated into software for analysis of data. It is expected that this exercise will stimulate further collaborative studies, which will help advance the capability of small-angle scattering to allow increasingly demanding experiments to be successfully performed.

**Figure 1**  (a) Atomic force microscope and (b) scanning electron microscope images of the PS3 polystyrene latex. The scale bars are 4 μm and 2 μm respectively.

**Figure 2**  Light scattering results for the sample with mass fraction $3 \times 10^{-6}$. (a) Static light scattering results shown as a Guinier plot of the natural logarithm of intensity versus $Q^2$. The gradient of the straight line is equal to $R_g^2/3$. The red squares indicate the deviation (multiplied by 100) between the fitted line and the measured data. (b) Dynamic light scattering data measured at a scattering angle of 90° with the fitted correlation function.

**Figure 3**  SANS data for PS3 Sample A from various instruments plotted (for each instrument the data is offset by a factor of 10). Data for SANS2D is not scaled.

**Figure 4**  Data from Figure 3 shown on expanded scales (a) $\log_{10} I$ vs. $Q$ and (b) linear scales. Although the logarithmic plot can give an overview, the significant systematic differences are seen most clearly in the linear plot.

**Figure 5**   Fits to the SANS data for the three samples, A, B and C (intensity increases with concentration) measured with D22 at the ILL using a sample to detector distance of 17.6 m and a 12 Å wavelength beam. The model incorporates a simple algorithm that allows for double scattering (Ghosh and Rennie, 2012) as well as instrument resolution and polydispersity. The particle radius was fitted as 724 Å with a Gaussian size distribution with standard deviation of 20 Å.

**Figure 6**   The influence of different factors on the scattering can be seen in the various fits to data for sample A. In (a) a fit that includes double scattering, polydispersity and resolution is compared to the calculation when the effect of double scattering is ignored but with sample parameters and resolution unchanged. The fit shown in (b) ignored multiple scattering but the polydispersity was allowed to vary to improve the agreement with the data. It is clearly seen that although the first minimum is not adequately smeared in this model, the large $Q$ data displays sharper minima than the model.

**Figure 7**   Plot of the ratio of the intensity for samples C and A measured with a 12 Å beam on D22. The measured data are compared to a Monte Carlo simulation. This representation of data is sensitive to differences in scattering from samples that occur with changes in concentration such as those that arise from interactions between particles, or for this case, from multiple scattering.

**Figure 8**   Small-angle X-ray scattering data from the polystyrene latex as (a) ) $\log_{10} I$ versus $\log_{10} Q$ plot and (b) Porod plot ($IQ^4$ versus $Q$).

**Table 1**   Samples used for SANS measurements

| Sample | Concentration / wt fraction | Volume fraction D$_2$O* | Scattering length density solvent / $10^{-6}$ Å$^{-2}$ |
|---|---|---|---|
| Sample A | 0.0039 | 0.95 | 6.00 |
| Sample B | 0.0010 | 0.98 | 6.21 |
| Sample C | 0.0003 | 0.995 | 6.32 |

*The remainder is H$_2$O

**Table 2**   SANS instruments used for measurements

| Instrument | Facility | λ / Å | Sample-Detector distance / m | Δλ/λ % | Reference |
|---|---|---|---|---|---|
| D22 | ILL | 6 & 12 | 17.6 | 10 | Cicognani (2008) |
| D11 | ILL | 6 & 13 | 8 & 34 | 9 | Lindner & Schweins (2010) |
| QUOKKA | ANSTO | 5.08 | 20 | 14 | Gilbert et al. (2006) |
| SANS2D | ISIS | 2.2 to 12.5 | 12 | * | Heenan et al. (2011) |
| NG7 | NCNR | 6 & 12 | 13.5 | 15 | Glinka et al. (1998) |

* Equivalent Δλ/λ for SANS2D for these measurements was about 3% at λ ~ 12Å for small Q, the beam geometry then dominates Q resolution, and about 17% at highest Q with λ ~ 2 Å where it was dominated by the 0.75 ms data collection time bin width that could have been considerably better, though the counts were low at this limit.

**Acknowledgements**   This activity was undertaken as part of the canSAS initiative on standardisation (http://www.cansas.org/wgwiki/index.php/Standardization_Working_Group).  ARR and MSH are grateful to the Swedish Research Council for partial support of this work.  The provision of beam time and the support of this project by the ILL, ISIS, NIST and Bragg Institute neutron facilities and the Diamond Light Source are gratefully acknowledged.  PDB and LP thank Dr Christopher Garvey for assistance with the QUOKKA instrument.  Use for some of the analysis of the SASview software is also acknowledged.  The mention of commercial products does not imply endorsement by NIST or other agencies, nor does it imply that the materials, software or equipment identified are necessarily the best available for the purpose.

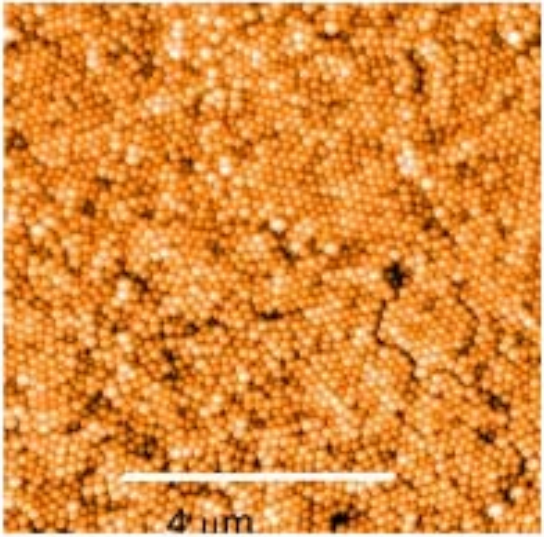 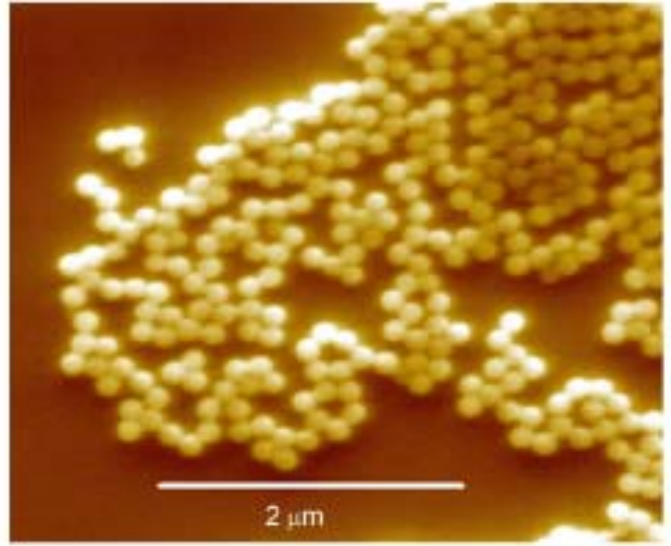

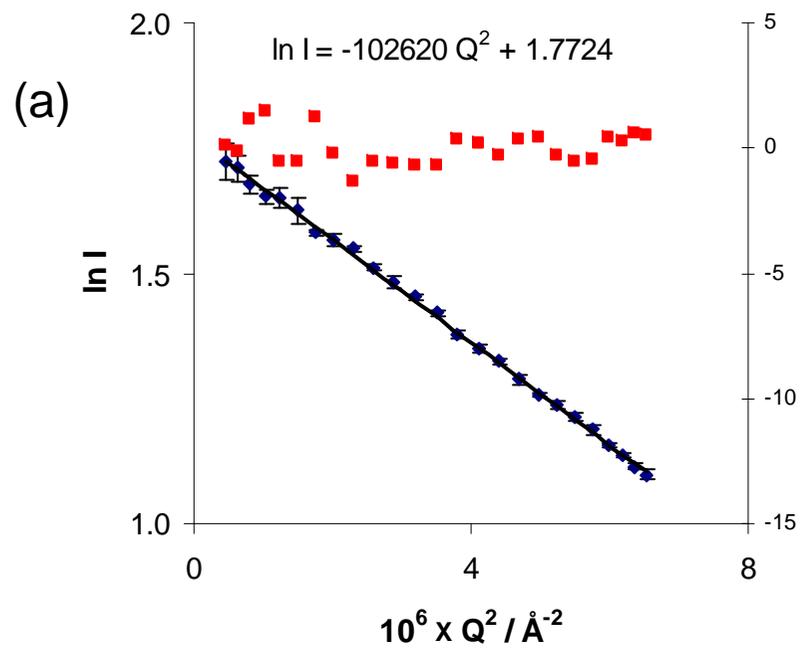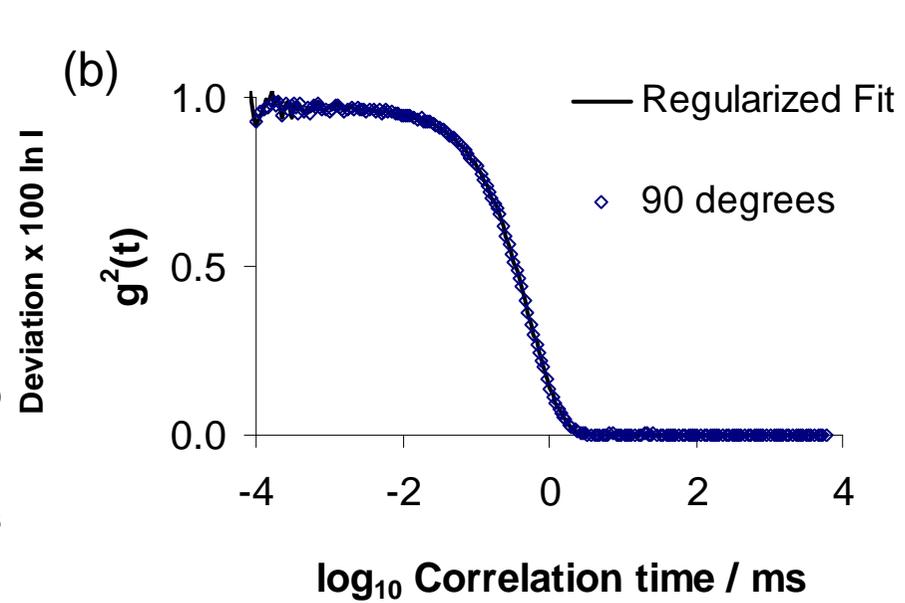

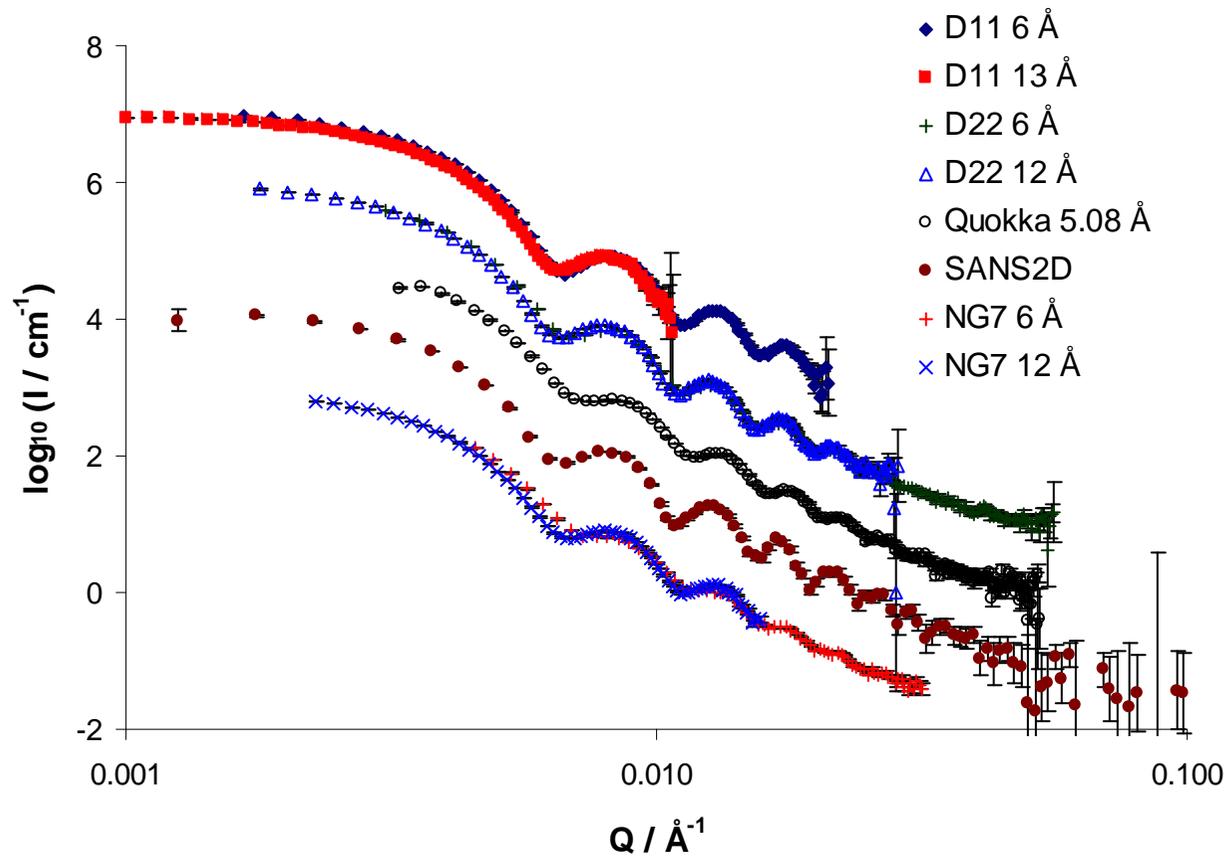

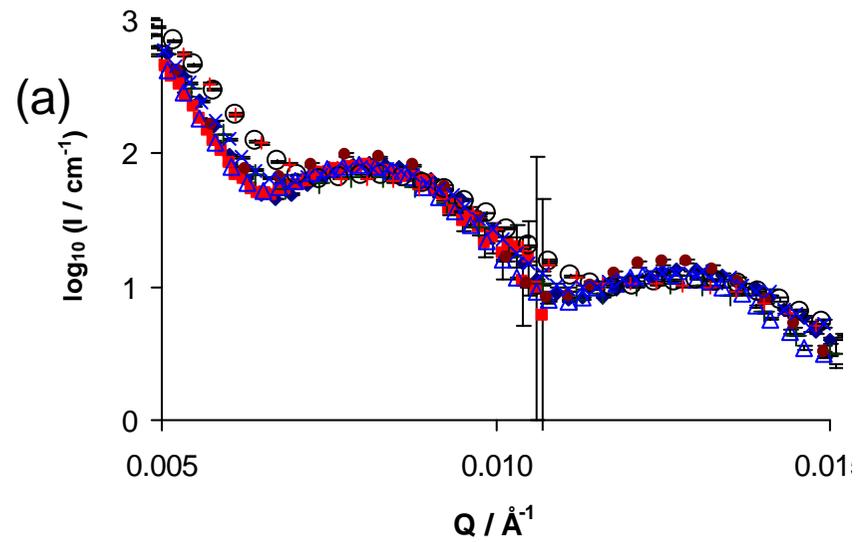 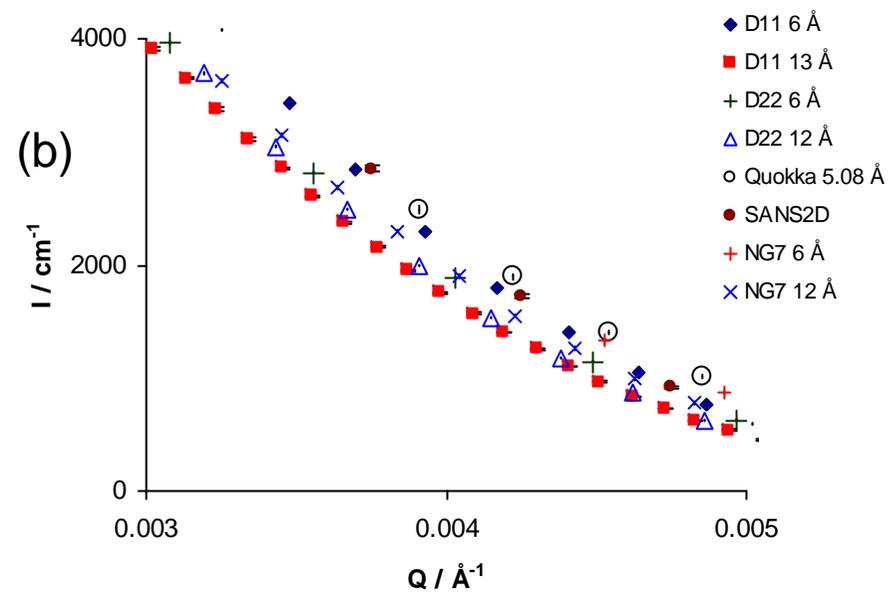

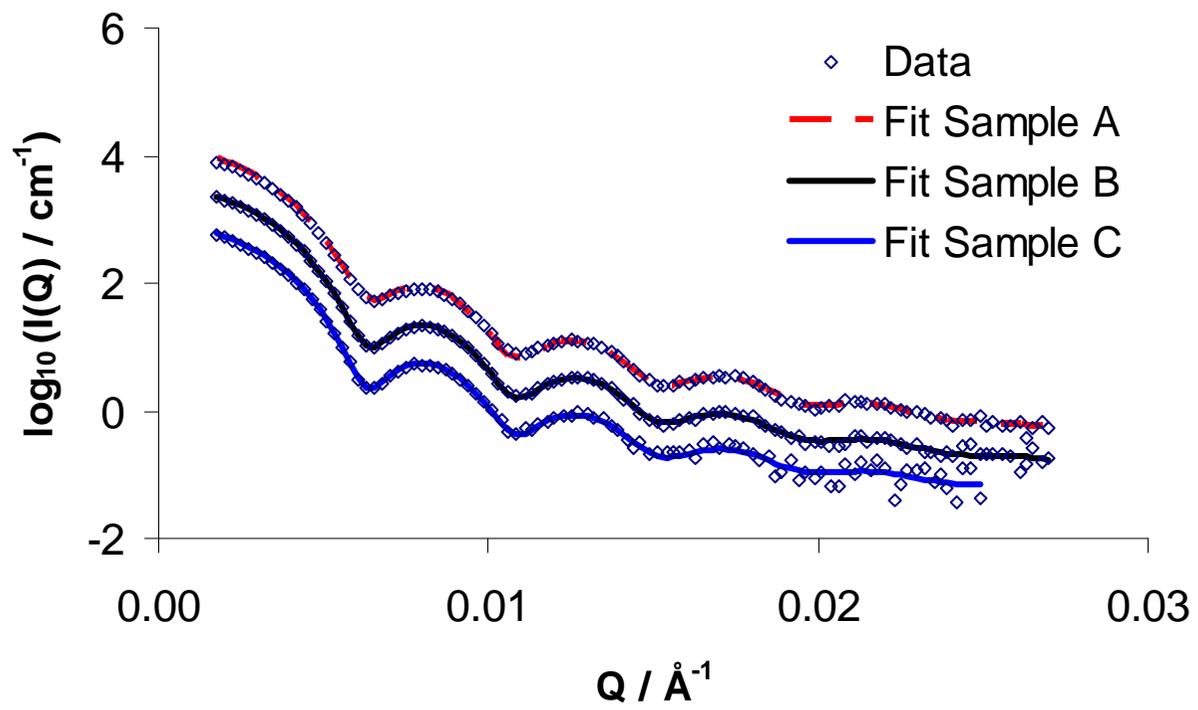

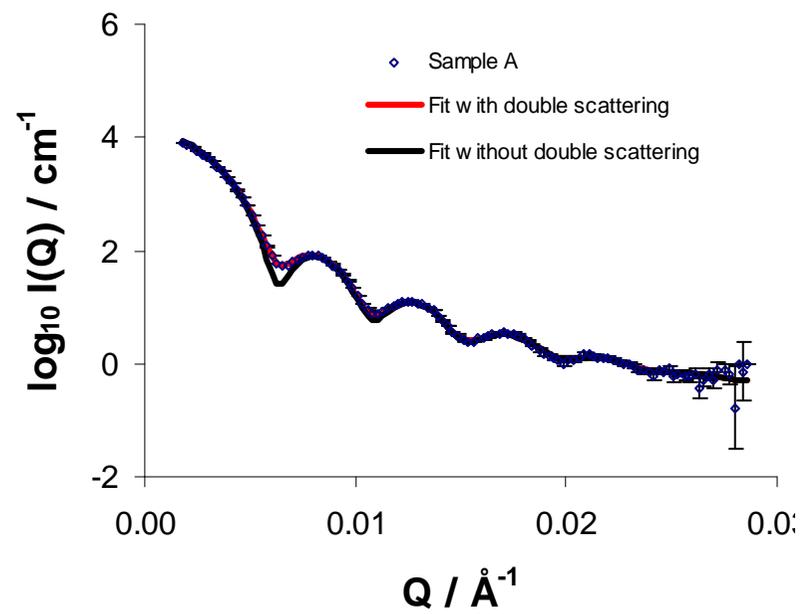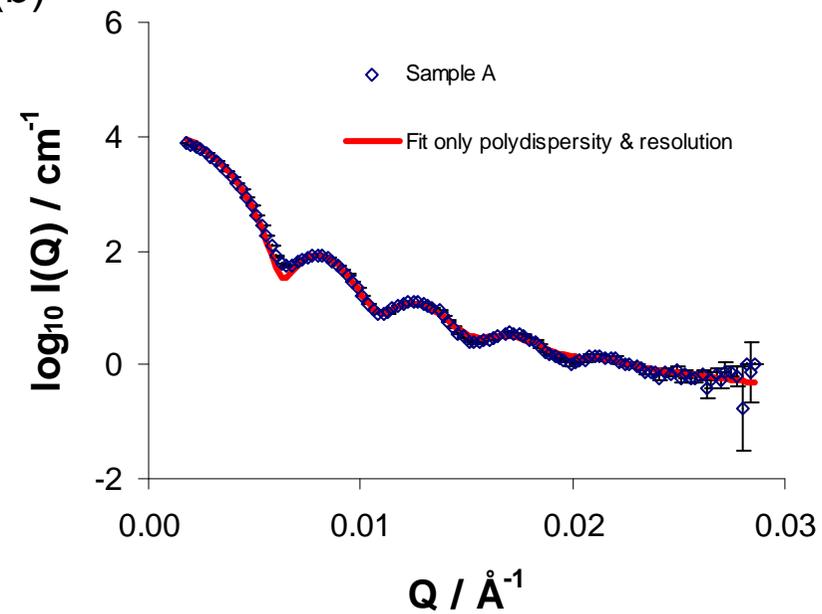

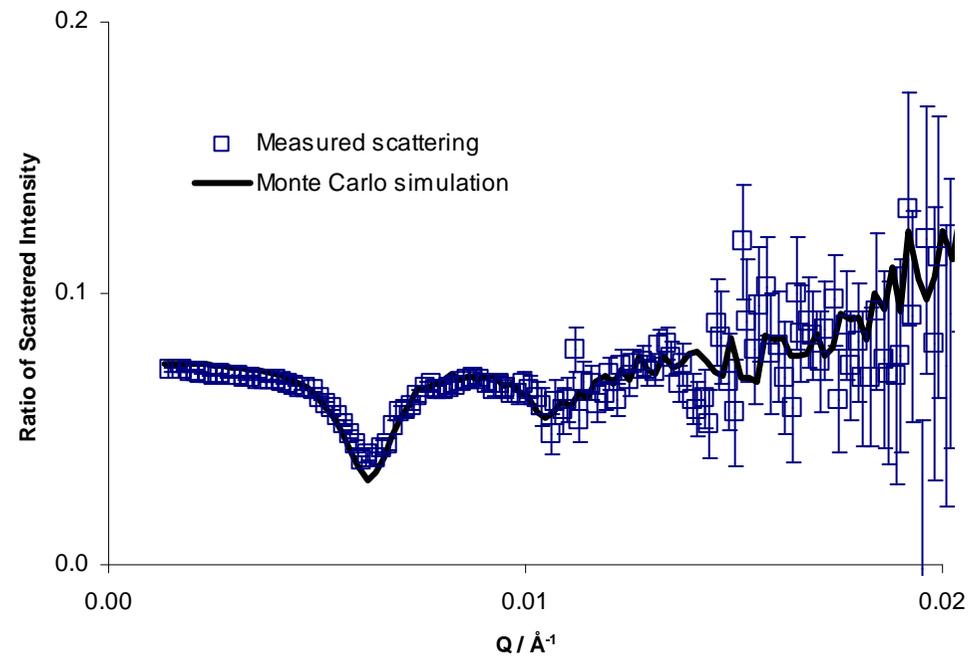

(a) 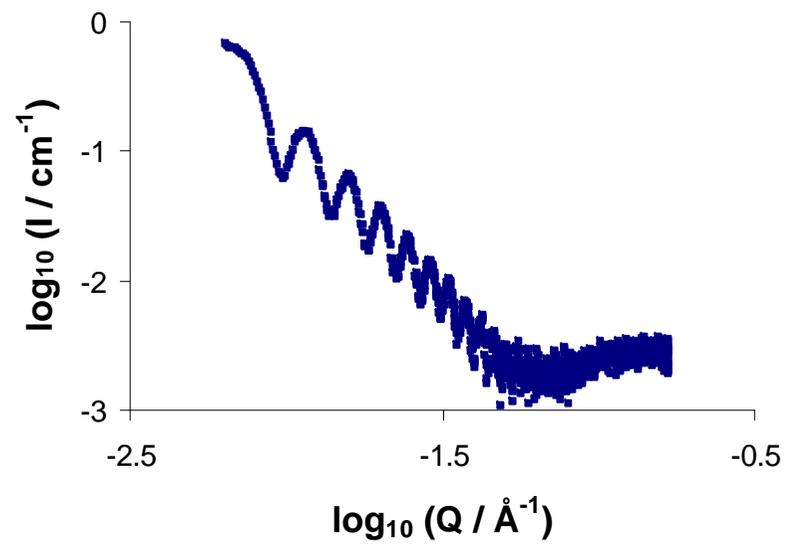

(b) 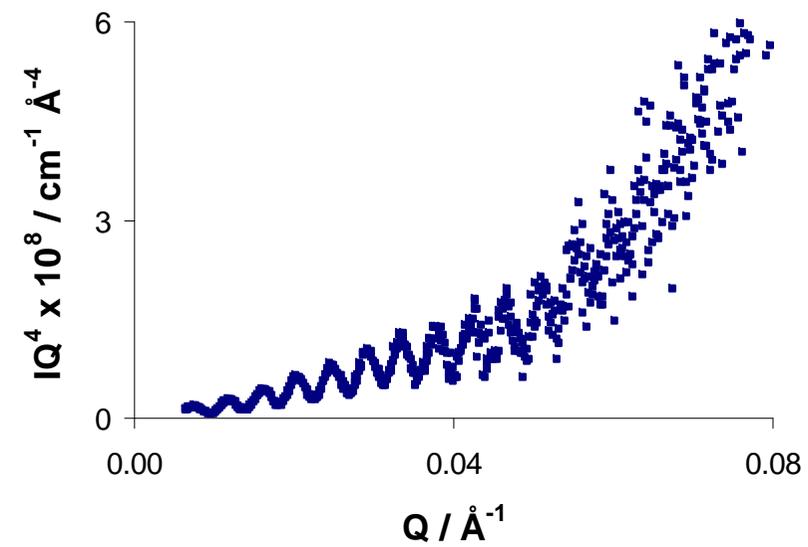